\documentclass[12pt,english]{article}
\usepackage{graphicx}
\usepackage{geometry}
\usepackage{color}
\usepackage{amsmath}
\usepackage{amssymb}
\usepackage{setspace}

\usepackage{multirow,multicol}
\usepackage[colorlinks,citecolor=blue,urlcolor=blue]{hyperref}
\usepackage[authoryear,round,colon]{natbib}
\usepackage{ulem}

\makeatletter
\@ifundefined{date}{}{\date{}}
\usepackage{latexsym}
\usepackage{bm}

\DeclareMathOperator*{\argmin}{arg\,min}

\newcommand{\bA}{\mathbf{A}}

\newcommand{\bx}{\bm{x}}
\newcommand{\by}{\bm{y}}

\newcommand{\bI}{\mathbf{I}}

\newcommand{\bP}{\mathbf{P}}
\newcommand{\bB}{\mathbf{B}}

\newcommand{\bD}{\mathbf{D}}
\newcommand{\bM}{\mathbf{M}}
\newcommand{\bX}{\mathbf{X}}
\newcommand{\bY}{\mathbf{Y}}
\newcommand{\bU}{\mathbf{U}}
\newcommand{\bV}{\mathbf{V}}

\newcommand{\bR}{\mathbf{R}}
\newcommand{\bS}{\mathbf{S}}

\newcommand{\bphi}{\boldsymbol{\phi}}
\newcommand{\bbeta}{\boldsymbol{\beta}}

\newcommand{\bgamma}{\boldsymbol{\gamma}}
\newcommand{\bepsilon}{\boldsymbol{\epsilon}}

\newcommand{\bSigma}{\boldsymbol{\Sigma}}

\newcommand{\bzero}{\mathbf{0}}

\def\E{E}

\usepackage{mathabx}

\newcommand{\norm}[1]{\Vert#1\Vert}
\newcommand{\Norm}[1]{\left\Vert#1\right\Vert}

\def\wh{\widehat}
\def\mbR{\mathbb R}
\def\mcC{\mathcal{C}}
\def\mcP{\mathcal{P}}
\def\mcN{\mathcal{N}}
\def\trans{^{\rm T}}
\newcommand{\bSx}{\bS_{\bX}}

\newtheorem{theorem}{Theorem}

\newcommand*{\indep}{%
  \mathbin{%
    \mathpalette{\@indep}{}%
  }%
}
\newcommand*{\nindep}{%
  \mathbin{
    \mathpalette{\@indep}{\not}
  }%
}
\newcommand*{\@indep}[2]{%
  \sbox0{$#1\perp\m@th$}
  \sbox2{$#1=$}
  \sbox4{$#1\vcenter{}$}
  \rlap{\copy0}
  \dimen@=\dimexpr\ht2-\ht4-.2pt\relax
  \kern\dimen@
  {#2}%
  \kern\dimen@
  \copy0 
} 

\makeatother

\title{Matrix Completion for Survey Data Prediction with Multivariate Missingness}
\author{Xiaojun Mao\thanks{School of Data Science, Fudan University, Shanghai
200433, P.R.C. Email: maoxj@fudan.edu.cn},\quad Zhonglei Wang\thanks{MOE Key Laboratory of Econometrics, Wang Yanan Institute for Studies in Economics and School of Economics, Xiamen University, Xiamen, Fujian 361005, P.R.C.
 Email: wangzl@xmu.edu.cn} \quad and \quad Shu Yang\thanks{Department of Statistics, North Carolina State University, North Carolina
27695, U.S.A. Email: syang24@ncsu.edu}}

\begin{document}


\date{}
\maketitle
\doublespacing


\label{firstpage}

\begin{abstract}
The  National Health and Nutrition Examination Survey (NHANES) studies the nutritional and health status over the whole U.S. population with comprehensive physical examinations and questionnaires.
However, survey data analyses become challenging due to inevitable missingness in almost all variables.
In this paper, we develop a new imputation method to deal with  multivariate missingness at random using matrix completion. In contrast to existing imputation schemes either conducting row-wise or column-wise imputation, we treat the data matrix as a whole which allows  exploiting both row and column patterns to impute the missing values in the whole data matrix at one time. 
We adopt a column-space-decomposition model for the population data matrix with easy-to-obtain demographic data as covariates and a low-rank structured residual matrix. 
A unique challenge arises due to lack of identification of parameters in the sample data matrix. 
We propose a projection strategy to uniquely identify the parameters and corresponding penalized estimators, which are computationally efficient and possess desired statistical properties.    
The simulation study shows that the doubly robust estimator using the proposed matrix completion for imputation has smaller mean squared error than other competitors. To demonstrate practical relevance, we apply the proposed method to the 2015-2016 NHANES Questionnaire Data.

\end{abstract}


\section{Introduction}

Survey data are the gold-standard for estimating finite population parameters and providing a comprehensive overview of the finite population at a given time. 
The National Health and Nutrition Examination Survey (NHANES, \texttt{https://www.cdc.gov/nchs/nhanes}), for example, is a program of studies to assess the health and nutrition status of the adults and children in the United States. The survey combines physical examinations and questionnaires and therefore can be used to provide a thorough and detailed health status assessment. 
In the 2015-2016 Questionnaire Data,
there are about 39 blocks of questions, such as dietary behavior and alcohol use,  and each block contains about ten relevant questions.

However, survey data analyses become challenging due to  inevitable multivariate missingness, leading to complex  “swiss cheese” patterns.
This occurs due to item nonresponse, when individuals provide answers
to partial but not all questions. 
Moreover, the missingness rates vary across questions and are extremely low for sensitive questions such as income. In the 
the NHANES 2015-2016 Questionnaire Data,
the average and standard error of the missingness rates are about 0.62 and 0.38, respectively.
This phenomenon is not an exception but a rule for
large surveys in the United States, including the  American National Election Studies, American Housing Survey and Current Population Survey. Inference ignoring the nonresponse items may be questionable \citep{rubin1976inference}

Imputation is widely used to handle item nonresponse, and existing methods for multivariate missingness can be categorized into two types: row-wise
imputation and column-wise imputation. For example, multiple imputation \citep{rubin1976inference,clogg1991multiple,fayproc,meng1994multiple,wang1998large,nielsen2003proper,kim2006bias,kim2011parametric,yang2016note}
can be viewed as a row-wise imputation method, which models the joint distribution
of all variables and generates the imputations based on a posterior
predictive distribution of the nonresponse items given the observed ones.
However, multiple imputation is sensitive to model misspecification,
especially when there are a lot of questions subject to non-response.
Moreover, it is computationally intensive, and it quickly becomes infeasible to implement as the number of questions subject to missingness increases. 
On the other hand, hot deck imputation \citep{chen2000nearest,kim2004fractional,fuller2005hot,andridge2010review} can be viewed as a column-wise imputation method. For subject
$i$ with missing $y_{ij}$ of the $j$th question, hot deck imputation methods search among
the units with responses to the $j$th question (referred to as donors
for the $j$th question), and impute the missing $y_{ij}$ by the response from its
nearest neighbor based on a certain distance
metric.

In contrast to most existing methods that use either models or  distance, 
we treat the data matrix as a whole and propose using matrix completion (\citealp{Candes-Recht09}; \citealp*{Keshavan-Montanari-Oh09}; \citealp*{Mazumder-Hastie-Tibshirani10}; \citealp*{Koltchinskii-Lounici-Tsybakov11}; \citealp{Negahban-Wainwright12}; \citealp{Cai-Zhou16}; \citealp{Robin-Klopp-Josse19}) as a tool for imputaiton, 
which allows exploiting both row and column patterns to impute the missing values in the whole data matrix at one time. 
Because there exist variables that are fully observed, 
we adopt a column-space-decomposition model \citep*{Mao-Chen-Wong19} for the population data matrix with easy-to-obtained demographic data as covariates and a low-rank structured residual matrix. 
The low-rank structure is due to underlying clusters of individuals and blocks of questions \citep{Candes-Recht09,Linden-Hambleton13,Davenport-Romberg16,Robin-Klopp-Josse19}.
Most works in the  matrix completion literature assume uniform missingness (or equivalently missingness completely at random), which however is unlikely to hold in the survey data context.   
Following \cite{Mao-Wong-Chen19}, we assume that the missing data mechanism is missingness at random (MAR; \citealp{rubin1976inference}).
Even though the population risk function identifies the parameter and data matrix uniquely, the sample risk function lacks identification in general. We propose a projection strategy so that   the new set of parameters can be identified after projection based on the sample data. For estimation, we consider  a risk function  weighted by both design weights and inverse of the estimated response probabilities, with the nuclear norm to encourage the low-rankness and two Frobenious norms to improve numerical performance of the penalized estimators.
After imputation for the sample data, we  use a doubly robust estimator for the population means \citep{kott1994note,bang2005doubly,kimhaziza,haziza2006nonresponse,kang2007demystifying,kott2010using}, which  is unbiased when the response model is correctly specified.

The proposed  method achieves the following advantages. First, it is computationally easy. Based on the column-space-decomposition model, we  have  modified  the objective function so that we can obtain a closed-form solution to recover the sample data matrix, and only one singular value decomposition (SVD) of an $n\times L$ matrix is required for computation. Second, it is a multi-purpose imputation method; that is, a single-imputation system  can be applied to all the survey questions. This is particularly attractive for a comprehensive analysis of the whole survey data.
Third, comparing to  fully parametric methods, we only require a low-rank assumption without any further specification. For theoretical investigation, we provide regularity conditions and  the asymptotic bounds of the penalized estimators and the doubly robust estimator.

The paper proceeds as follows. Section~\ref{sec:Basic-setup} provides
the basic setup and estimation procedure of the proposed method. Section~\ref{sec: asymptotic properties} discusses the theoretical properties of the proposed method. 
A simulation study
is conducted in Section~\ref{sec: simulation} to illustrate
the advantage of the proposed method compared with other  competitors. Section~\ref{sec: application} 
presents an application to the NHANES 2015-2016 Questionnaire Data. Some concluding remarks are given in Section~ \ref{sec:conclusion}. 

\section{Basic Setup\label{sec:Basic-setup}}
\subsection{Notation, Assumption and Model}
Consider a finite population of $N$ subjects with study variables $U_N=\{(\bx_i,\by_i):i=1,\ldots,N\}$.
Organize the finite population  in matrix forms $\bX_N=(x_{ij})\in \mathbb{R}^{N\times d}$ and
$\bY_N = (y_{ij})\in \mathbb{R}^{N\times L}$, where $\bX_N$ is fully observed, and $\bY_N$ is subject to missingness. 
We are interested in estimating  $\theta_j = N^{-1}\sum_{i=1}^Ny_{ij}$ for $j=1,\ldots,L$.

Assume that the finite population is a realization of an infinite population, called a \textit{super-population}, and consider a super-population model $\zeta$, 
\begin{equation}
\bY_N = \bA_N + \bepsilon_N,\label{eq: csd model 1}
\end{equation}
where $\bA_N\in \mathbb{R}^{N\times L}$ represents the structural component of the data matrix, and
$\bepsilon_N = (\epsilon_{ij}) \in \mathbb{R}^{N\times L}$ is a matrix of independent errors with $E(\epsilon_{ij})=0$ for $i=1,\ldots,N$ and $j=1,\ldots,L$. 
Following \citet{Candes-Recht09} ,\citet{Linden-Hambleton13}, \citet{Davenport-Romberg16} and \citet{Robin-Klopp-Josse19}, we assume that $\bA_N$ has a low-rank structure, which is reasonable in the survey context. On the one hand, the finite population can be divided into groups by demographics such as age, gender, address and occupations. On the other hand, survey questions can also be grouped into several blocks. For example, in the NHANES 2015-1016 Questionnaire Data, there exist different blocks of questions, such as health and nutrition status, education, income level and so on, and each block contains several relevant questions.

To further incorporate $\bX_N$ into the model (\ref{eq: csd model 1}), following \citet{Mao-Chen-Wong19},  we adopt the column-space-decomposition model,
\begin{equation}
\bA_N =\bX_N\bbeta^{\ast}+\bB_N^{\ast},
\label{eq: csd model}
\end{equation}
where $\bbeta^{\ast}=(\beta_{ij})$ is a $d\times L$ coefficient matrix, and $\bB_N^{\ast}$ is an $N\times L$ low-rank matrix, inherited from the low-rank structure of $\bA_N$.
To avoid identification issues, we assume that $\bX_N\trans\bB_N^{\ast}=\bzero$. We also assume that the elements in the matrices are indexed by $N$ implicitly.  

Following \cite{fay1991design}
and \cite{shao1999variance}, we first have a census with nonrespondents. 
Denote $\bR_N=(r_{ij})\in \mathbb{R}^{N\times L}$ as the response indicator matrix with $r_{ij}=1$ if $y_{ij}$ is observed and $r_{ij}=0$ otherwise. Following most of the missing data literature, we assume that the missing data mechanism is MAR. Specifically, assume that $\bX_N$ explains the missing  mechanism well in the sense that the values in $\by_i$ are MAR conditional on $\bx_i$. 
Under MAR, the response probability becomes $p_{ij}=\Pr(r_{ij}=1\mid \bX_N,\bY_N)=\Pr(r_{ij}=1\mid \bx_i)$. 
For regularity reasons, we require $p_{ij}$ to be bounded away from 0 and 1 for all $i$ and $j$. 
Following most of the empirical literature, we assume  that the response probability follows a logistic regression model, 
\begin{equation}\label{eqn:logit}
p_{ij}=p_{ij}\left(\bx_{i}\right) = \frac{\exp\left\{\left(1,\bx_{i}\trans\right)\bgamma_{.j}\right\}}{1+\exp\left\{\left(1,\bx_{i}\trans\right)\bgamma_{.j}\right\}},
\end{equation}
where $\bgamma_{.j}\in\mbR^{d+1}$  is the parameter vector specific for the $j$th column of $\bY_N$.
Denote $\bP_N^{\dagger}=(p_{ij}^{-1})\in \mathbb{R}^{N\times L}$ as the matrix of the inverse response probabilities. 

To motivate the proposed method, we first consider the population data. Denote $\mcC(\bX)$ as the column space of a matrix $\bX$ and $\mcN(\bX)=\{\bM\in\mathbb{R}^{N\times L}:\bX\trans\bM=\bzero\}$. Under Model (\ref{eq: csd model}) and MAR, for any $\bbeta\in\mbR^{d\times L}$ and $\bB\in\mcN(\bX_N)$, the population risk function $R(\bbeta,\bB)$ is
\begin{equation}
R(\bbeta,\bB)=\frac{1}{NL}\E\Norm{\bR_N\circ\bP_N^{\dagger}\circ\bY_N - \bX_N\bbeta-\bB}_{F}^{2}, \label{eq: R population missing}
\end{equation}
where ``$\circ$'' is the Hardamard product, and $\norm{\bM}_F = (\sum_{i=1}^N\sum_{j=1}^Lm_{ij}^2)^{1/2}$ is the Frobenius norm of an $N\times L$ matrix $\bM=(m_{ij})$. Then, $(\bbeta^{\ast},\bB_N^{\ast})$ in (\ref{eq: csd model}) uniquely minimizes the population risk function $R(\bbeta,\bB)$; see \citet{Mao-Chen-Wong19} for details. 

In practice, it is both time-consuming and expensive to conduct a census for a finite population. 
Survey sampling has been the gold standard to estimate finite population parameters based on a relatively small probability sample.
Assume that a sample of size $n$ is selected by a probability sampling design \citep{fuller2009sampling}. 
Denote $I_i$ as the sampling indicator; specifically,  $I_i=1$ if the $i$th subject is sampled and 0 otherwise. 
Let $\pi_i = E(I_i \mid U_N)$ be the inclusion probability of the $i$th subject, where the expectation is taken with respect to the probability sampling mechanism. For example, Poisson sampling generates a sample using $N$ independent Bernoulli trials, where $I_i$ is generated from a Bernoulli distribution with success probability $\pi_i$ for $i=1,\ldots,N$.
Without loss of generality, assume that the first $n$ subjects of the finite population are sampled.  
In the following, without ambiguity, we use $\bM_n$ to denote the sample data matrix of the first $n$ rows of a population data matrix $\bM_N$.  
If $\bY_n$ is fully observed, we can obtain a Horvitz-Thompson  estimator \citep{Horvitz1952} of $\theta_j$, 
\begin{equation}
\wh\theta_j = \frac{1}{N}\sum_{i=1}^N\frac{I_i}{\pi_{i}}y_{ij}.\label{eq: HT estimator}
\end{equation}
It follows that $\wh\theta_j$ is a design-unbiased estimator of $\theta_j$, that is, $E(\wh\theta_j\mid U_N) =\theta_j$.

In the presence of missingness in $\bY_n$, we propose to use matrix completion as an imputation method to recover the missing values. With the sample data matrices,  we use the following empirical risk function to approximate the population risk $R(\bbeta,\bB)$ in (\ref{eq: R population missing}):
\begin{eqnarray}
R^{\ast}(\bbeta,\bB_n)&=&\frac{1}{NL}\sum_{i=1}^{N}\frac{I_{i}}{\pi_{i}}\sum_{j=1}^{L}\left\{\frac{r_{ij}}{p_{ij}}y_{ij}-(\bX_N\bbeta)_{ij}-b_{ij}\right\}^2 \label{eq: risk function}\\
&=&\frac{1}{NL}\Norm{ \bD_n^{-1/2}\left(\bR_n\circ \bP_n^{\dagger}\circ \bY_n-\bX_n\bbeta-\bB_n\right)}_F^{2}, \label{eq: risk function 2}
\end{eqnarray}
where $\bD_n=\mbox{diag}(\pi_1,\ldots,\pi_n)$ is a diagonal matrix with $\pi_i$ being the $(i,i)$th entry. If the sampling mechanism is non-informative, there is no need to adjust sampling weights for estimating $\bbeta^{\ast}$ and $\bB_N^{\ast}$ in (\ref{eq: csd model}). Adjusting for sampling weights, however, achieves two goals. First, the expectation of (\ref{eq: risk function}) is the population risk function $R(\bbeta,\bB)$, so we target for estimating the population data matrix instead of the sample data matrix. Second, it allows for informative sampling. Under informative sampling, the empirical risk function without sampling weights is biased of the population risk function. 

\subsection{Non-identifiability of \texorpdfstring{$(\boldsymbol{\beta}^{\ast},\bB_n^{\ast})$}{}\label{sec:nonidentifiability}}

In the population  risk function (\ref{eq: R population missing}), $\bX_N\trans\bB_N^{\ast}=\bzero$ guarantees that $(\bbeta^{\ast},\bB_N^{\ast})$ is identifiable; see \citet{Mao-Chen-Wong19}. Moreover, the decomposition of $\bA_N$ into $\bX_N\bbeta^{\ast}\in \mathcal{C}(\bX_N)$ and $\bB_N^{\ast}\in \mathcal{N}(\bX_N)$
gives  benefits for showing theoretical properties of the estimators and encourages an efficient algorithm allowing for a closed-form solution of $\wh{\bB}_N$. However,  the same decomposition technique may fail  to guarantee identification of parameters in the sample risk function $R^{\ast}(\bbeta,\bB_n)$ in (\ref{eq: risk function 2}) because  $(\bD_{n}^{-1/2}\bX_n)\trans(\bD_{n}^{-1/2}\bB_n)=\bX_n\trans\bD_{n}^{-1}\bB_n$ may not be a zero matrix. Even for simple random sampling with $\pi_i=n/N$  for $i=1,\ldots,N$, we cannot ensure $\bX_n\trans\bD_{n}^{-1}\bB_n=Nn^{-1}\bX_n\trans\bB_n=\bzero$. It means that there is no space restriction for both $\bbeta$ and $\bB_n$ in $R^{\ast}(\bbeta,\bB_n)$. Thus, for any $(\bbeta,\bB_n)$ and nonzero $\bbeta_1$, we always have $R^{\ast}(\bbeta,\bB_n)=R^{\ast}(\bbeta+\bbeta_1,\bB_n-\bX_n\bbeta_1)$.

To deal with the lack of identifiability, we consider a decomposition of  $\bD_n^{-1/2}(\bR_n\circ \bP_n^{\dagger}\circ \bY_n-\bX_n\bbeta-\bB_n)$ by
\[
\bD_n^{-1/2}\left(\bR_n\circ \bP_n^{\dagger}\circ \bY_n\right)-\bD_n^{-1/2}\bX_n\bbeta-\mcP_{\bD_n^{-1/2}\bX_n}(\bD_n^{-1/2}\bB_n)-\mcP_{\bD_n^{-1/2}\bX_n}^{\perp}(\bD_n^{-1/2}\bB_n).
\]
where 
$\mcP_{\bD_n^{-1/2}\bX_n}=\bD_n^{-1/2}\bX_n(\bX_n\trans\bD_n^{-1}\bX_n)^{-1}\bX_n\trans\bD_n^{-1/2}$, $\mcP_{\bD_n^{-1/2}\bX_n}^{\perp} = \bI -  \mcP_{\bD_n^{-1/2}\bX_n}$ and $\bI$  is the $n\times n$ identity matrix. Denote 
\[
\bbeta^{\ast\prime}=\bbeta^{\ast}+(\bX_n\trans\bD_n^{-1}\bX_n)^{-1}\bX_n\trans\bD_n^{-1}\bB_n^{\ast}\qquad\text{and}\qquad\bB_n^{\ast\prime}=\mcP_{\bD_n^{-1/2}\bX_n}^{\perp}(\bD_n^{-1/2}\bB_n^{\ast}),
\]
respectively. Then, we have $\bB_n^{\ast\prime}\in\mcN(\bD_n^{-1/2}\bX_n)$, so we can decompose the objective function $R^{\ast}(\bbeta,\bB_n)$ as
\begin{align*}
R^{\ast}(\bbeta,\bB_n) = R^{\ast}(\bbeta^{\prime},\bB_n^{\prime})=&\frac{1}{NL}\left[\Norm{\mcP_{\bD_n^{-1/2}\bX_n}\left\{\bD_n^{-1/2}\left(\bR_n\circ \bP_n^{\dagger}\circ \bY_n\right)\right\}-\bD_n^{-1/2}\bX_n\bbeta^{\prime}}_F^{2}+\right.\\
&\left.\Norm{\mcP_{\bD_n^{-1/2}\bX_n}^{\perp}\left\{\bD_n^{-1/2}\left(\bR_n\circ \bP_n^{\dagger}\circ \bY_n\right)\right\}-\bB_n^{\prime}}_F^{2}\right].
\end{align*}
It can be seen  that $\bbeta^{\ast\prime}$ and $\bB_n^{\ast\prime}$ are the unique minimizers of $R^{\ast}(\bbeta^{\prime},\bB_n^{\prime})$. Although $\bbeta^{\ast}$ and $\bB_n^{\ast}$ cannot be uniquely determined,  we ensure that
\[
\bX_n\bbeta^{\ast\prime}+\bD_n^{1/2}\bB_n^{\ast\prime}=\bX_n\bbeta^{\ast}+\bB_n^{\ast},
\]
which is sufficient to identify 
the parameters of interest $\theta_j$ for $j=1,\ldots,L$. Therefore, in what follows, we will focus on estimating $\bbeta^{\ast\prime}$ and $\bB_n^{\ast\prime}$.

\subsection{Estimation of \texorpdfstring{$\boldsymbol{\beta}^{\ast\prime}$ and $\mathbf{B}_n^{\ast\prime}$}{}}

Because $\bP_n$ is unknown, 
we consider a maximum likelihood estimator $\wh\bP_n$ of $\bP_n$ and 
\begin{align*}
\wh R^{\ast}(\bbeta^{\prime},\bB_n^{\prime})=&\frac{1}{NL}\left[\Norm{\mcP_{\bD_n^{-1/2}\bX_n}\left\{\bD_n^{-1/2}\left(\bR_n\circ \wh\bP_n^{\dagger}\circ \bY_n\right)\right\}-\bD_n^{-1/2}\bX_n\bbeta^{\prime}}_F^{2}+\right.\\
&\left.\Norm{\mcP_{\bD_n^{-1/2}\bX_n}^{\perp}\left\{\bD_n^{-1/2}\left(\bR_n\circ \wh\bP_n^{\dagger}\circ \bY_n\right)\right\}-\bB_n^{\prime}}_F^{2}\right],
\end{align*}
where $\wh\bP_n^{\dagger}$ is the matrix of the estimated response probabilities.
Since $\bbeta^{\prime}$ and $\bB_n^{\prime}$ are high-dimensional parameters, a direct minimization of $\wh R^{\ast}(\bbeta,\bB_n)$ would often result in over-fitting. To avoid such an issue, we incorporate penalty terms for those two parameters. Specifically, we propose the penalized estimators of $(\bbeta^{\ast\prime},\bB_{n}^{\ast\prime})$ as
\begin{equation}\label{eqn:minbetaB}
(\wh\bbeta^{\prime},\wh\bB_{n}^{\prime})=\argmin_{\bbeta^{\prime},\bB_n^{\prime}\in\mcN(\bD_n^{-1/2}\bX_n)}\wh R^{\ast}(\bbeta^{\prime},\bB_n^{\prime}) +\tau_1\Norm{\bbeta^{\prime}}_{F}^2+ \tau_2\left\{\alpha\Norm{\bB_n^{\prime}}_{\ast}+(1-\alpha)\Norm{\bB_n^{\prime}}_{F}^2\right\},
\end{equation}
where $\norm{\bM}_{\ast}=\mbox{trace}(\sqrt{\bM\trans\bM})$ is the nuclear norm of a real-valued matrix $\bM$, and $\tau_1$, $\tau_2>0$ along with $0\le\alpha\le1$ are regularization parameters. 
Since $\bB_N^{\ast}$ is assumed to be low-rank, $\bB_n^{\ast}$ is also low-rank and  $\text{rank}(\bB_{n}^{\ast\prime}) = \text{rank}(\bB_n^{\ast})$. 
Similar to the rank sum norm, the nuclear norm also encourages a low-rank solution. 
In matrix completion literature, one reason why  people consider the   nuclear norm instead of the rank norm penalty directly is that the minimization problem with rank norm penalty is NP-hard \citep{Candes-Recht09}. The two additional Frobenius norm terms of $\bbeta^{\prime}$ and $\bB_n^{\prime}$ are applied to improve finite sample performance \citep{Zou-Hastie05,Sun-Zhang12,Mao-Chen-Wong19}.

To obtain $\wh\bbeta^{\prime}$, it is essentially a solution of a ridge regression problem, and we have
\begin{equation*}\label{eqn:closedbeta}
\wh\bbeta^{\prime}=\left(\bX_n\trans\bD_n^{-1}\bX_n+NL\tau_1\bI\right)^{-1}\bX_n\trans\bD_n^{-1}\left(\bR_n\circ\wh\bP_n^{\dagger}\circ\bY_n\right).
\end{equation*}
To obtain  $\wh\bB_n^{\prime}$, following the same argument in Proposition~2 of \citet{Mao-Chen-Wong19}, we can extend the searching domain for $\bB_n^{\prime}\in\mcN(\bD_n^{-1/2}\bX_n)$ in the minimization problem \eqref{eqn:minbetaB} to be $\bB_n^{\prime}\in\mbR^{n\times L}$. This allows us to express the solution $\wh\bB_n^{\prime}$ in a closed form. Let $\bU\bSigma \bV\trans$ be the SVD of a matrix $\bM$, where $\bSigma = \mathrm{diag}(\{\sigma_{i}\})$.
Define the corresponding singular value soft-thresholding operator $\mathcal{T}_c$ by
\begin{equation*}
\mathcal{T}_c \left(\bM\right) =\bU \mathrm{diag}(\{ \left(\sigma_i-c\right)_{+} \})\bV^\intercal
\label{eqn:T}
\end{equation*}
for any $c\ge 0$,
where $x_{+}=\max(x,0)$.
It can be shown that the solution $\wh\bB_n^{\prime}$ in \eqref{eqn:minbetaB} possesses the following closed form:
\begin{equation*}
\wh\bB_n^{\prime} =
\frac{1}{1+\left(1-\alpha\right)NL\tau_2}
\mathcal{T}_{\alpha NL\tau_2/2}
\left[\mcP_{\bD_n^{-1/2}\bX_n}^{\perp}\left\{\bD_n^{-1/2}\left(\bR_n\circ\wh\bP_n^{\dagger}\circ\bY_n\right)\right\}\right].\label{eqn:hatBB}
\end{equation*}
Following the common practice in  matrix completion works (\citealp{Mazumder-Hastie-Tibshirani10}; \citealp*{Xu-Jin-Zhou13}; \citealp*{Chiang-Hsieh-Dhillon15}; \citealp*{Mao-Wong-Chen19}), we obtain tuning parameters $\tau_1$, $\tau_2$ and $\alpha$ by a $5$-fold cross validation procedure. After obtaining $(\wh\bbeta^{\prime},\wh\bB_n^{\prime})$, an estimator of $\bA_n$ is given by 
\begin{equation}
\wh\bA_n=\bX_n\wh\bbeta^{\prime}+\bD_n^{1/2}\wh\bB_n^{\prime}.\label{eq: estimate A}
\end{equation}

\subsection{Comparison with Hot Deck Imputation and Multiple Imputation\label{sec:comp}}

It is worth comparing the proposed matrix completion method with existing approaches for imputation.
Hot deck imputation 
uses an observed datum as a ``donor'' to impute each missing item based on a  specific distance using some fully observed auxiliary information. For hot deck imputation, an underlying regression model, $f_j(\bx_{i})$, is assumed for the item $y_{ij}$.
Therefore, only $\bx_i$ is used for imputing $y_{ij}$ but not $y_{ik}$ with $k\neq j$.
The multiple imputation \citep{rubin1978multiple}
assumes a joint model of $(\bx_i,\by_i)$ and uses all available variables for imputation. However, fully parametric modeling is sensitive to model misspecification.

In our approach, the low-rank structure of $\bA_N$ suggests a general decomposition of $\bA_N$ to be $\bA_N=\bU_N\bV_N\trans$, where $\bU_N\in\mbR^{N\times r_{\bA_N}}$ and $\bV_N\in\mbR^{L\times r_{\bA_N}}$ are two hidden matrices. Due to the low-rank assumption, we have $r_{\bA_N}\ll N$ and $r_{\bA_N}\ll L$. In our column-space-decomposition model, we enforce part of the hidden matrix $\bU_N$ to be a fully observed matrix $\bX_N\in\mbR^{N\times d}$ and denote the corresponding part in $\bV_N$ to be $\bbeta^{\ast}$, where $\bbeta^{\ast}$ is just a different notation and still totally unknown. Thus, the decomposition could be written as $\bA_N=(\bX_N,\bU_N^{\ast})(\bbeta^{\ast},\bV_N^{\ast})\trans$ with $\bB_N^{\ast} = \bU_N^{\ast}{\bV_N^{\ast}}\trans$.   In a general setting, the only restriction for $\bU_N^{\ast}$ is $\text{rank}(\bX_N,\bU_N^{\ast})=r_{\bA_N}$, which means that each column of $\bU_N^{\ast}$ cannot be fully expressed by the columns in $\bX_N$. However, it still allows for $\text{cor}(\bX_N,\bU_N^{\ast})\neq \bzero$. Then, it is difficulty to identify the hidden matrix $\bU_N^{\ast}$ under the general setting. Thus, we restrict the column space of $\bU_N^{\ast}$ to be orthogonal to the column space of $\bX_N$. Fortunately,  the number of covariates $d$ is usually fixed and $d\ll r_{\bA_N}$, which means that we would not lose too much freedom for $\bU_N^{\ast}$.

\subsection{Estimation of $\theta_j$}

After imputation, it may be natural to estimate $\theta_j$ by the Horvitz-Thompson estimator (\ref{eq: HT estimator}) applied to the imputed dataset. However, it is well known that the estimated low-rank matrix $\wh\bB_n^{\prime}$ is biased when $n$ is finite \citep{Mazumder-Hastie-Tibshirani10,Foucart-Needell-Plan17,Carpentier-Kim18,Chen-Fan-Ma19}. Therefore, the resulting imputation estimator is biased.   
Researchers have proposed different procedures to alleviate or eliminate the bias. 
\citet{Mazumder-Hastie-Tibshirani10} suggested a post-processing step  by re-estimating the  estimated singular values without any theoretical guarantee.
\citet{Foucart-Needell-Plan17} proposed an algorithm based on projection onto the max-norm ball to de-bias the estimator under non-uniform and deterministic sampling patterns. 
\citet{Carpentier-Kim18} considered an estimator using an iterative hard thresholding method and showed that the entry-wise bias is small when the sampling design is Gaussian. 
More recently, \citet{Chen-Fan-Ma19} developed a de-biasing procedure using a similar idea to de-biasing LASSO estimators and showed nearly optimal properties for the resulting estimator. 
Despite these advances in literature,  the scenarios considered above are restricted to deterministic sampling, Gaussian sampling or missing completely at random (MCAR), which are not applicable in our setting.

We use a simple strategy borrowing the idea from the doubly robust estimation literature \citep*{robins1994estimation,bang2005doubly,cao2009improving} and consider a doubly robust estimator of $\theta_j$ as
\begin{equation}
\wh{\theta}_{j,DR} = \frac{1}{N}\sum_{i=1}^{N}\frac{I_i}{\pi_i}\left\{\frac{r_{ij}(y_{ij}-\wh{a}_{ij})}{\wh{p}_{ij}}+\wh{a}_{ij}\right\},\label{eq: double robust estimator}
\end{equation}
where $\wh{p}_{ij}$ and $\wh{a}_{ij}$  are the $(i,j)$th element of $\wh{P}_n$ and $\wh{A}_n$, respectively.
It can be shown that 
\begin{equation*}
\wh{\theta}_{j,DR} = \frac{1}{N}\sum_{i=1}^{N}\frac{I_i}{\pi_i}\left\{\frac{r_{ij}(y_{ij}-\wh{a}_{ij})}{{p}_{ij}}+\wh{a}_{ij}\right\}+o_P(1)
\end{equation*}
when the response model (\ref{eqn:logit}) is correctly specified, so $\wh{\theta}_{j,DR}$ is asymptotically unbiased for $\theta_j$ for this case.

\section{Asymptotic Properties\label{sec: asymptotic properties}}

In this section, we first study the asymptotic properties of the estimator $\wh\bA_n$ in (\ref{eq: estimate A}) under the logistic regression model \eqref{eqn:logit}. Further, we establish the average convergence rate of $\wh\theta_{j,DR}-\theta_j$ for $j=1,\ldots,L$.

For asymptotic inference, we follow the framework of \citet{isaki1982survey} and assume that both the population size $N$ and the sample size $n$ go to infinity. 
Let $\norm{\bM}=\sigma_{\max}(\bM)$ and $\norm{\bM}_{\infty}=\max_{i,j}\vert m_{ij}\vert$ be the spectral and the maximum norms of a matrix $\bM$, respectively. We use the symbol ``$\asymp$'' to represent the asymptotic equivalence in order, that is, $a_n\asymp b_n$ is equivalent to $a_n=O(b_n)$ and $b_n=O(a_n)$.

The technical conditions needed for our analysis are given as follows.
\begin{enumerate}
	\item[C1] (a) The random errors $\{\epsilon_{ij}\}_{i,j=1}^{N,L}$ in   \eqref{eq: csd model} are independently distributed random variables such that $\E(\epsilon_{ij})=0$ and $\E(\epsilon^2_{ij})=\sigma_{ij}^2<\infty$ for all $i,j$. (b) For some finite positive constants $c_{\sigma}$ and $\eta$, 
	$\underset{i,j}{\max}\,\E|\epsilon_{ij}|^{l}\le\frac{1}{2}l!c_{\sigma}^{2}\eta^{l-2}$ for any positive integer $l\ge 2$.
	\item[C2] The inclusion probability satisfies  $\pi_i\asymp nN^{-1}$ for $i=1,\ldots,N$.
	\item[C3] The population design matrix $\bX_N$ is of size $N\times d$ such that $N>d$. Moreover, there exists a positive constant $a_{x}$ such that $\norm{\bX_N}_{\infty}\le a_{x}$ and $\bX_N\trans\bD_N\bX_N$ is invertible, where $\bD_N$ is a diagonal matrix with $\pi_i$ as its $(i,i)$th entry. Furthermore, there exists a symmetric matrix $\bSx$ with $\sigma_{\min}(\bSx)\asymp1\asymp\norm{\bSx}$ such that $n_0^{-1}\bX_N\trans\bD_N\bX_N\to\bSx$ as $N\to\infty$, where $n_0=\sum_{i=1}^N\pi_i$ is the expected sample.
	\item[C4] There exists a positive constant $a$ such that $\max\{\norm{\bX_N\bbeta^{\ast}}_{\infty},\norm{\bA_N}_{\infty}\}\le a$.
	\item[C5]The indicators of observed entries $\{r_{ij}\}_{i,j=1}^{N,L}$ are mutually independent, $r_{ij}\sim\text{Bern}(p_{ij})$ for $p_{ij}\in (0,1)$ and are independent of $\{\epsilon_{ij}\}_{i,j=1}^{N,L}$ given $\bX_N$. Furthermore, for $i=1,\dots,N$ and $j=1,\dots,L$, $\Pr(r_{ij}=1 | \bx_{i}, y_{ij}) = \Pr(r_{ij}=1 | \bx_{i})$  follows the logistic regression model \eqref{eqn:logit}.
	\item[C6] There exists a lower bound $p_{\min}\in (0,1)$ such that $\underset{i,j}{\min}\{p_{ij}\}\ge p_{\min}>0$, where $p_{\min}$ is allowed to depend on $n$ and $L$. The number of questions $L\le n$.
\end{enumerate}

Condition~C1(a) is a common regularity condition for the measurement errors in $\bepsilon_{N}$, and C1(b) is the Bernstein condition \citep{Koltchinskii-Lounici-Tsybakov11}. Condition~C2 is widely used in survey sampling and regulates the inclusion probabilities of a sampling design \citep{fuller2009sampling}. To illustrate ideas, we consider Poisson sampling in this section, and our discussion applies to other sampling designs such as simple random sampling and probability-proportional-to-size sampling. In Condition~C3, the requirement $N>d$ is easily met as the number of questions in a survey is usually fixed, and the population size is often larger than the number of questions. As the dimension of $n_0^{-1}\bX_N\trans\bD_N\bX_N$ is fixed at $d\times d$, it is mild to assume $\bX_N\trans\bD_N\bX_N$ to be invertible, and there exists a symmetric matrix $\bSx$ as the probability limit of $n_0^{-1}\bX_N\trans\bD_N\bX_N$. Furthermore, the sample size is often larger than the number of questions, that is, $n>d$, and it is not hard to show that together with Condition~C2, the probability limit of $n^{-1}\bX_n\trans\bX_n$ is also $\bSx$ under Poisson sampling; see the Supplementary Materials \citep*{Mao-Wang-Yang} for details. The order of $\sigma_{\min}(\bSx)$ and $\norm{\bSx}$ equals to $1$ is due to $\norm{\bX_N}_{\infty}<\infty$.
Condition~C4 is also standard in the matrix completion literature \citep{Koltchinskii-Lounici-Tsybakov11,Negahban-Wainwright12,Cai-Zhou16}. Especially, it is reasonable to assume all the responses are bounded in survey sampling.  Condition~C5 describes the independent Bernoulli model for the response indicator of observing $y_{ij}$, where the probability of observation $p_{ij}$ follows the logistic model \eqref{eqn:logit}. In Condition~C6, 
the lower bound $p_{\min}$ is allowed to go to $0$ with $n$ and $L$ growing. This condition is more general than we need for a typical survey, and $p_{\min}\asymp 1$ suffices. Typically, the number of questions $L$ grows slower than the number of participants $n$ in survey sampling. Thus, the assumption that $L\le n$ is quite mild.

For any $\delta_{\sigma}>0$, some positive constants $C_d$, $C_g$, $C$ and $t \in (d+3,+\infty)$, define
\begin{equation}\label{eqn:Deltans}
\Delta\left(\delta_{\sigma},t\right) = \max\left\{N^{1/2}n^{-1}L^{-1}\log^{1/2}\left(n\right)p_{\min}^{-1/2},N^{1/2}n^{-5/4}L^{-1/4}\log^{1/2}\left(L\right)\log^{\delta_{\sigma}/4}\left(n\right)t^{1/2}p_{\min}^{-3/2}\right\},
\end{equation}
and $\eta_{n,L}(\delta_{\sigma},t) = 4/(n+L)+ 4C_{d}t\exp\{-t/2\}+4/L+C\log^{-\delta_{\sigma}}(n)$. We can verify that $\lim_{t\to\infty}\{\lim_{n,L\to\infty}\eta_{n,L}(\delta_{\sigma},t)\}=0$. Once we have  $n^{1/2}L^{-3/2}\log\left(n\right)p_{\min}^{2}\ge (d+3)$, by choosing $t$ such that
\begin{equation}\label{eqn:tlogistic}
d+3<t<n^{1/2}L^{-3/2}\log\left(n\right)p_{\min}^{2},
\end{equation}
we can show $\underset{t}{\sup}\Delta(\delta_{\sigma},t)\asymp N^{1/2}n^{-1}L^{-1}\log^{1/2}(n)p_{\min}^{-1/2}$, which is denoted by $\Delta(\delta_{\sigma})$. Here, the requirement $n^{1/2}L^{-3/2}\log\left(n\right)p_{\min}^{2}\ge (d+3)$ is easy to fulfill once $n$ large enough.

\begin{theorem}\label{thm:logistic}
	Assume Conditions C1-C6 and Poisson sampling, $p_{\min}^{-1}=O(L\log^{-1}(n+L))$ and 
	the logistic model \eqref{eqn:logit} hold. Choose $t$ as \eqref{eqn:tlogistic}, $\tau_1\asymp N^{-1}n L^{-1}{\log^{-1/2}(n)}\Delta(\delta_{\sigma})$, $1-\alpha\asymp (nL)^{-1}$, $\tau_2\asymp p_{\min}^{-3/2}N^{-1}n^{1/4}L^{-1/4}\log^{1/2}(L)\log^{\delta_{\sigma}/3}(n)$ in \eqref{eqn:minbetaB} for any $\delta_{\sigma}>0$. Then, for some positive
	constant $C_1$ and $C_2$, with probability at least $1-\eta_{n,L}(\delta_{\sigma},t)$, we have 
	\begin{align*}
	\quad \frac{1}{mL}\Norm{\wh\bbeta^{\prime}-\bbeta^{\ast\prime}}_F^{2}\le
	C_1r_{\bB_N} L^{-1}\log\left(n\right)p_{\min}^{-1}\quad \text{and}\\
	\quad \frac{1}{nL}\Norm{\wh\bB_n^{\prime}-\bB_n^{\ast\prime}}_F^{2}\le
	C_2r_{\bB_N} Nn^{-1}L^{-1}\log\left(n\right)p_{\min}^{-1}.
	\end{align*}
\end{theorem}

A proof of Theorem~\ref{thm:logistic} is given in the Supplementary Materials \citep{Mao-Wang-Yang}. As $\lim_{t\to\infty}\{\lim_{n,L\to\infty}\eta_{n,L}(\delta_{\sigma},t)\}=0$, Theorem~\ref{thm:logistic} implies that 
$(mL)^{-1}\norm{\wh\bbeta^{\prime}- \bbeta^{\ast\prime}}_{F}^{2}=O_{p}\{r_{\bB_N} L^{-1}\log(n)p_{\min}^{-1}\}$ and $(nL)^{-1}\norm{\wh\bB_n^{\prime}- \bB_n^{\ast\prime}}_{F}^{2}=O_{p}\{r_{\bB_N} Nn^{-1}L^{-1}\log(n)p_{\min}^{-1}\}$.

As we pointed out in Section~\ref{sec:nonidentifiability}, even with the knowledge of $(\bbeta^{\ast\prime},\bB_n^{\ast\prime})$, we cannot recover $(\bbeta^{\ast},\bB_n^{\ast})$ exactly. Fortunately, we have 
\[
\wh\bA_n=\bX_n\wh\bbeta^{\prime}+\bD_n^{1/2}\wh\bB_n^{\prime},
\]
which enables us to derive the asymptotic bound for $(nL)^{-1}\norm{\wh\bA_n-\bA_n}_{F}^{2}$ given in the following theorem.

\begin{theorem}\label{thm:A}
	Assume that the same conditions in Theorem~\ref{thm:logistic} hold. For a positive
	constant $C_3$, with probability at least $1-\eta_{n,L}(\delta_{\sigma},t)$, we have
	\begin{align*}
	\frac{1}{nL}\Norm{\wh\bA_n-\bA_n}_F^{2}\le
	C_3r_{\bB_N}L^{-1}\log\left(n\right)p_{\min}^{-1}.
	\end{align*}
\end{theorem}

A brief proof of Theorem~\ref{thm:A} can be found in the Supplementary Materials. The term $(nL)^{-1}\norm{\wh\bA_n-\bA_n}_{F}^{2}$ has the same order with upper bound of $(mL)^{-1}\norm{\wh\bbeta^{\prime}-\bbeta^{\ast\prime}}_{F}^{2}$. To ensure the convergence of $(nL)^{-1}\norm{\wh\bA_n-\bA_n}_{F}^{2}$, we only require that $n=O\{\exp(r_{\bB_N}^{-1}Lp_{\min})\}$ which is quite mild. In survey sampling, it is reasonable to assume that $p_{\min}\asymp1$, especially when the participants are awarded. Thus, the assumption that $p_{\min}^{-1}=O(L\log^{-1}(n+L))$ is easy to fulfill once $L$ large enough. It can be shown that the convergence rate for $(nL)^{-1}\norm{\wh\bA_n-\bA_n}_{F}^{2}$ can be simplified to $r_{\bB_N}L^{-1}\log(n)$ if $p_{\min}\asymp1$. As we have discussed in Section~\ref{sec:comp}, the proposed method  achieves robustness against model misspecification. 

The following theorem provides the average convergence rate of $\wh\theta_{j,DR}$ for $j=1,\ldots,L$. 
\begin{theorem}\label{thm: thetadr }
	Assume that the same conditions in Theorem~\ref{thm:logistic} hold and $p_{\min}\asymp1$. Then, we have $$
	\frac{1}{L}\sum_{j=1}^L(\wh{\theta}_{j,DR} - \theta_j)^2 =O_p\{r_{\bB_N}L^{-1}\log\left(n\right)\}.
	$$
\end{theorem}
A proof for Theorem~\ref{thm: thetadr } is given in the Supplementary Materials. By Theorem~\ref{thm: thetadr }, the mean squared difference between $\wh{\theta}_{j,DR}$ and $\theta_j$ among the $L$ questions is bounded by $O_p\{r_{\bB_N}L^{-1}\log\left(n\right)\}$. To ensure the convergence of $L^{-1}\sum_{j=1}^L(\wh{\theta}_{j,DR} - \theta_j)^2$, similarly with before, we only require that $n=O\{\exp(r_{\bB_N}^{-1}L)\}$ which is quite mild.

\section{Simulation\label{sec: simulation}} 
We use (\ref{eq: csd model 1}) and (\ref{eq: csd model}) to generate a finite population $U_N$, where elements in $\bX_N$ and $\bbeta^{\ast}$ are generated by $ \mcN(0.5,1^2)$, $\bB_N^{\ast} = \mcP_{\bX_N}^{\perp}\bB_L\bB_R$, $\bB_L$ is an $N\times k$ matrix, $\bB_R$ is a $k\times L$ matrix, elements of $\bB_L$ and $\bB_R$ are generated by $ \mcN(1,3^2)$, elements of $\bepsilon_N$ are generated such that the signal-noise ratio is 2, $N=10\,000$ is the population size, $L=500$ is the number of questions in the survey, $d=20$ is the rank of $\bX_N$ and $\bbeta^{\ast}$,  and $k=10$ is the rank of $\bB_N^{\ast}$. From the  generated finite population $U_N$,  the following sampling designs are considered:
\begin{enumerate}
	\item[I] Poisson sampling with inclusion probability $\pi_i = ns_i(\sum_{i=1}^Ns_i)^{-1}$, where $s_i>0$ is a size measure of the $i$th subject, and the generation of $s_i$ is discussed later. Specifically, for  $i=1,\ldots,N$, a sampling indicator $I_i$ is generated by a Bernoulli distribution with success probability $\pi_i$.
	\item[II] Simple random sampling with sample size $n$.
	\item[III] Probability-proportional-to-size sampling with size measure $s_i$. That is, a sample of size $n$ is selected independently from the finite population $U_N$ with replacement, and the selection probability of the $i$th subject is  proportional to its size measure $s_i$.
\end{enumerate}
We consider two scenarios for the sampling procedure. One is informative sampling with $s_i = 7^{-1}\sum_{j=1}^7y_{ij}-m_{s}+1$, where $m_s = \min\{7^{-1}\sum_{j=1}^7y_{ij}:i=1,\ldots,N\}$. The other is noninformative sampling with $s_i = d^{-1}\sum_{j=1}^dx_{ij}+e_i+1$, where $e_i\sim\mbox{Ex}(1)$, and $\mbox{Ex}(\lambda)$ is an exponential distribution with rate parameter $\lambda$. Two different sample sizes are considered, $n=200$ and $n=500$, and the following estimation methods are compared:
\begin{enumerate}
	\renewcommand{\labelenumi}{\Roman{enumi}}
	\item Hot deck imputation. For each item with $r_{ij}=0$, we use $y_{kj}$ as the imputed value, where $\bx_k$ is nearest to $\bx_j$ among  $\{\bx_l:r_{lj}=1\}$. Treating the imputed values as observed ones, we   estimate $\theta_j$ by (\ref{eq: HT estimator}).
	\item Multiple imputation. We adopt the multivariate imputation by chained equations (MICE) by \citet{buuren2010mice}. MICE fully specifies the conditional distribution for the missing data and uses a posterior predictive distribution to generate imputed values for the nonresponse items; check \citet{buuren2010mice} for details. However, it is impossible for MICE to impute all missing responses in $\bY_n$ at the same time due to the computational issues. For comparison, we only use the first 20 items of $\bY_n$ to specify the conditional distribution for MICE and generate imputed values for the corresponding nonresponses. Then, we can use (\ref{eq: HT estimator}) to estimate $\theta_j$.
	\item Inverse probability method. For $j=1,\ldots,L$,  a logistic regression model (\ref{eqn:logit}) is fitted.
	Then, $\theta_j$ is estimated by 
	$$
	\wh\theta_{j,IPM} = N^{-1}\sum_{i=1}^nr_{ij}\wh{p}_{ij}^{-1}y_{ij}.
	$$
	\item Doubly robust estimator using linear regression model. For $j=1,\ldots,L$, consider the following linear regression model:
	\begin{equation}
	y_{ij} = \phi_{0j} +\bx_i\trans\bphi_{1j},\label{eq: linear regression}
	\end{equation}
	and the parameters in (\ref{eq: linear regression}) are estimated by 
	$$
	(\wh\phi_{0j},\wh\bphi_{1j}) = \argmin_{(\phi_{0j},\bphi_{1j})}\sum_{i=1}^n\frac{r_{ij}}{\pi_i\wh{p}_{ij}}(y_{ij}-\phi_{0j} -\bx_i\trans\bphi_{1j})^2.
	$$
	Then, we can use a doubly robust estimator based on  the linear model  (\ref{eq: linear regression}) to estimate $\theta_j$.
	\item Doubly robust estimator using naive imputation. We use the naive imputation method \citep{Mazumder-Hastie-Tibshirani10} by assuming MCAR to generate the imputed values, and use the doubly robust estimator to estimate $\theta_j$.
	\item Doubly robust estimator using the proposed method in (\ref{eq: double robust estimator}).
\end{enumerate}
For comparison, we also consider the Horvitz-Thompson estimator in (\ref{eq: HT estimator}) using the fully observed data.

We conduct 1\,000 Monte Carlo simulations. 
Table \ref{tab: table 1} shows the Monte Carlo bias and standard error for the first five items under informative probability-proportional-to-size sampling with sample size $n=500$. Specifically, the Monte Carlo bias and standard error for the $j$th question are obtained by 
$$
\mbox{Bias}_j = \wh\theta_j^{(m)}-\theta_j \qquad\text{and}\qquad \mbox{SE}_j = \frac{1}{1\,000}\sum_{m=1}^{1\,000}(\wh\theta_j^{(m)}-\wh\theta_j)^2,
$$
respectively, where $\wh\theta_j = {1\,000^{-1}}\sum_{m=1}^{1\,000}\wh\theta_j^{(m)}$, and $\wh\theta_j^{(m)}$ is an estimator from a specific estimation method in the $m$th Monte Carlo simulation. The standard error for the hot deck imputation is  much larger compared with other methods. The bias of the multiple imputation is larger than the inverse probability method and doubly robust estimators since the model is misspecified.  Besides, the multiple imputation method is not preferable due to the computation complexity, especially when the number of items is large. Two doubly robust estimators using naive imputation and the proposed method have smaller variability than others, and the bias for the doubly robust estimator using the proposed method is smaller. Compared with the Horvitz-Thompson estimator using the fully observed data, the variance of the doubly robust estimator using the proposed method is larger.

\begin{table}[ht]
	\centering
	\caption{Monte Carlo bias (Bias) and standard error (SE) for the first five items under informative probability-proportional-to-size sampling with sample size $n=500$. ``HDI'' is the hot deck imputation,  ``IPM'' is the inverse probability method,  ``DRLR'' is the doubly robust estimator using linear regression, ``DRNI'' is the doubly robust estimator using naive imputation,  ``DRMC'' is the doubly robust estimator using the proposed method, and ``Full'' is the Horvitz-Thompson estimator using fully observed data.}\label{tab: table 1}
	\begin{tabular}{ccrrrrr}\hline\hline
		\multirow{2}{*}{Method} & \multirow{2}{*}{Stat.} & \multicolumn{5}{c}{Items}\\ 
		&  & I & II & III & IV & V  \\ 
		\hline\multirow{2}{*}{HDI} & Bias & -0.13 & 1.40 & 0.36 & 1.35 & 1.16 \\ 
		& SE & 8.03 & 12.20 & 13.19 & 12.75 & 9.54 \\ 
		&  &  &  &  &  &  \\ 
		\multirow{2}{*}{MI} & Bias & 0.29 & 0.64 & 0.30 & 0.69 & 0.56 \\ 
		& SE & 1.03 & 1.53 & 1.60 & 1.62 & 1.18 \\ 
		&  &  &  &  &  &  \\ 
		\multirow{2}{*}{IPM} & Bias & -0.03 & -0.02 & -0.01 & 0.05 & 0.14 \\ 
		& SE & 1.07 & 1.74 & 1.81 & 1.86 & 1.26 \\ 
		&  &  &  &  &  &  \\ 
		\multirow{2}{*}{DRLR} & Bias & 0.00 & 0.11 & 0.01 & 0.16 & 0.21 \\ 
		& SE & 1.07 & 1.76 & 1.82 & 1.88 & 1.26 \\ 
		&  &  &  &  &  &  \\ 
		\multirow{2}{*}{DRNI} & Bias & -0.26 & -0.25 & -0.53 & -0.51 & -0.38 \\ 
		& SE & 0.94 & 1.43 & 1.47 & 1.50 & 1.06 \\ 
		&  &  &  &  &  &  \\ 
		\multirow{2}{*}{DRMC} & Bias & -0.16 & 0.02 & -0.21 & -0.11 & -0.04 \\ 
		& SE & 0.94 & 1.38 & 1.43 & 1.48 & 1.03 \\ 
		&  &  &  &  &  &  \\ 
		\multirow{2}{*}{Full} & Bias & -0.04 & -0.01 & -0.00 & 0.01 & 0.06 \\ 
		& SE & 0.78 & 1.27 & 1.33 & 1.36 & 0.92 \\ 
		\hline
	\end{tabular}\bigskip
\end{table}

Next, we compare different estimation methods by the Monte Carlo mean squared error (MSE)
$$
\mbox{MSE}_j = \frac{1}{1\,000}\sum_{m=1}^{1\,000}(\wh\theta_j^{(m)}-\theta_j)^2\quad (j=1,\ldots,L).
$$
The result for multiple imputation is omitted due to the computational issue.  Table \ref{tab: MSE simulation} summarizes the mean and standard error of MSEs for different questions. From Table \ref{tab: MSE simulation}, we can conclude that the mean MSE and its standard error of the doubly robust estimator using the proposed method are smallest among alternatives for all scenarios. Besides, the average MSE and its standard error of doubly robust estimator using the proposed  method are slightly larger than the Horvitz-Thompson estimator using fully observed data.

\begin{table}
	\centering
	\caption{Summary of MSE for different estimation methods. ``NIF'' shows the results under noninformative sampling, and ``IF'' shows those under informative sampling.  ``POI'' for the Poisson sampling, ``SRS'' for the simple random sampling, and ``PPS'' stands for the probability-proportional-to-size sampling.  ``Mean'' and ``SE'' are the  mean and standard error of the MSEs for $L=500$ items, ``HDI'' is the hot deck imputation,  ``IPM'' is the inverse probability method,  ``DRLR'' is the doubly robust estimator using linear regression, ``DRNI'' is the doubly robust estimator using naive imputation,  ``DRMC'' is the doubly robust estimator using the proposed  matrix completion
		method, and ``Full'' is the Horvitz-Thompson estimator using fully observed data.  }\label{tab: MSE simulation}
	\begin{tabular}{ccccrrrrrr}
		\hline\hline
		&\multirow{2}{*}{Design} & \multirow{2}{*}{Sample size} &\multirow{2}{*}{Stat.}&\multicolumn{6}{c}{Estimation methods}\\
		&&& &HDI & IPM&DRLR &DRNI&   DRMC& Full\\
		\hline   
		\multirow{14}{*}{NIF}  & \multirow{4}{*}{POI} & \multirow{2}{*}{$n=200$} & Mean & 17.48 & 11.70 & 12.48 & 8.36 & 7.65 & 6.26 \\ 
		&  &  & SE & 8.41 & 4.30 & 4.70 & 2.94 & 2.48 & 2.26 \\ 
		&  & \multirow{2}{*}{$n=500$} & Mean & 15.34 & 4.54 & 4.63 & 3.45 & 2.92 & 2.45 \\ 
		&  &  & SE & 7.69 & 1.70 & 1.75 & 1.39 & 0.99 & 0.89 \\ \\
		& \multirow{4}{*}{SRS} & \multirow{2}{*}{$n=200$} & Mean & 16.91 & 9.97 & 10.91 & 7.28 & 6.36 & 5.11 \\ 
		&  &  & SE & 8.26 & 3.81 & 4.00 & 2.72 & 2.05 & 1.91 \\ 
		&  & \multirow{2}{*}{$n=500$} & Mean & 15.12 & 3.82 & 3.96 & 3.01 & 2.44 & 1.96 \\ 
		&  &  & SE & 7.79 & 1.46 & 1.49 & 1.23 & 0.81 & 0.76\\\\
		& \multirow{4}{*}{PPS} & \multirow{2}{*}{$n=200$} & Mean & 17.10 & 11.16 & 12.04 & 7.98 & 7.10 & 5.82 \\ 
		&  &  & SE & 8.32 & 4.29 & 4.56 & 2.93 & 2.35 & 2.10 \\ 
		&  & \multirow{2}{*}{$n=500$} & Mean & 15.32 & 4.47 & 4.57 & 3.37 & 2.85 & 2.37 \\ 
		&  &  & SE & 7.84 & 1.64 & 1.69 & 1.34 & 0.91 & 0.88 \\ \\
		\\
		\multirow{14}{*}{IF}   & \multirow{4}{*}{POI} & \multirow{2}{*}{$n=200$} & Mean & 17.25 & 11.23 & 12.07 & 8.11 & 6.98 & 5.97 \\ 
		&  &  & SE & 8.17 & 4.26 & 4.51 & 2.98 & 2.24 & 2.14 \\ 
		&  & \multirow{2}{*}{$n=500$} & Mean & 15.34 & 4.35 & 4.43 & 3.32 & 2.70 & 2.32 \\ 
		&  &  & SE & 7.70 & 1.68 & 1.70 & 1.38 & 0.84 & 0.81 \\ \\
		& \multirow{4}{*}{SRS} & \multirow{2}{*}{$n=200$} & Mean & 16.79 & 9.86 & 10.80 & 7.32 & 6.40 & 5.11 \\ 
		&  &  & SE & 8.32 & 3.74 & 4.10 & 2.83 & 2.11 & 1.90 \\ 
		&  & \multirow{2}{*}{$n=500$} & Mean & 14.99 & 3.83 & 3.91 & 2.96 & 2.41 & 1.97 \\ 
		&  &  & SE & 7.74 & 1.42 & 1.44 & 1.18 & 0.78 & 0.73 \\ \\
		& \multirow{4}{*}{PPS} & \multirow{2}{*}{$n=200$} & Mean & 16.91 & 10.64 & 11.65 & 7.66 & 6.54 & 5.45 \\ 
		&  &  & SE & 8.01 & 4.07 & 4.26 & 2.64 & 2.08 & 2.02 \\ 
		&  & \multirow{2}{*}{$n=500$} & Mean & 15.20 & 4.29 & 4.38 & 3.19 & 2.58 & 2.22 \\ 
		&  &  & SE & 7.70 & 1.60 & 1.61 & 1.24 & 0.82 & 0.81 \\ \\
		\hline
	\end{tabular}
	\bigskip
	
\end{table}

\section{Application\label{sec: application}}
The NHANES 2015-2016 Questionnaire Data is used as an application for the proposed method. Conducted by the  National Center for Health Statistics, the NHANES is a unique survey combining interviews and physical examinations to study the health and nutritional status of adults and children in the United States. Data are released in a two-year cycle. The sample size is approximately 5\,000, and the participants are nationally representative. The sampling design for NHANES aims at reliable estimation for  population subgroups formed by age, sex, income status and origins. Specifically, the Questionnaire Data contains family-level information including food security status as well as individual level information including dietary behavior and alcohol use.

In this section, we are interested in estimating the population mean of  alcohol usage, blood pressure and cholesterol, diet behavior and nutrition, diabetes status, mental health, income status, and sleep disorders based on the newly released NHANES 2015-2016 Questionnaire Data. There are about 39 blocks of questions in this dataset. Each block contains several relevant questions, and  the number of questions in our analysis ranges from 4 to 17. There are $n=5\,735$ eligible subjects involved and 146 items including 45 demographic questions. Among the demographic items, there are 16 fully observed items including age, gender and race-ethnicity, and they are used as the covariates $\bX_n$ in (\ref{eq: estimate A}). In addition, the sampling weight for each subject is available. For the questions in our study, the average and standard error of the missing rates are 0.33 and 0.37, respectively.

For estimating the population mean of each question, we consider those estimation methods in Section~\ref{sec: simulation}, and the covariates are standardized. Since the population size is unavailable, we use $\wh{N}=\sum_{i=1}^nw_i$ instead, where $w_i$ is the sampling weight of the $i$th subject incorporating the sampling design as well as calibration \citep{fuller2009sampling}.
For the multiple imputation, we only impute the missing values for the first 20 items due to the computational issue.

Table \ref{tab: application} shows missing rates and estimation results for six randomly selected items grouped by the missing rate.  There are two items with low missing rates 0.08 and 0.09, two with middle missing rates 0.26 and 0.29, and two with high missing rates 0.65 and 0.68. Besides, there are three items are among the first 20 used for the multiple imputation. Thus, the selected questions are representative.  When the missing rate is low, estimators are similar for different methods. As missing rate increases, estimators for the multiple imputation, hot deck imputation and the double robust estimator using naive matrix completion are different from those for the inverse probability method and double robust estimators using linear regression and the proposed method. When the missing rate is large, say around $.65$, the double robust estimator using linear regression differs from those for inverse probability method and the proposed method. 
Noting that all estimators are unbiased if the response model is corrected specified; however, the doubly robust estimator with matrix completion provide the most accurate estimation when all questions are of interest.

\begin{table}[ht]
	\centering
	\caption{Estimation results for six questions. ``I'' for ``Family has savings more than \$20,000''  ``II'' stands for ``Had at least 12 alcohol drinks/1 yr?'', ``III'' for ``How often drink alcohol over past 12 mos?'', ``IV'' for ``How often drank milk age 5-12?'', 
		``V'' for ``Told had high blood pressure - 2+ times?'', and ``VI'' for ``Receive community/Government meals delivered?'' 
	}\label{tab: application}
	\begin{tabular}{crrrrrrrr}\hline\hline
		\multirow{2}{*}{Items} & \multirow{2}{*}{Missing rate} &  &\multicolumn{6}{c}{Estimation methods}\\ 
		&  &  & MI & HDI & IPM & DRLR & DRNI & DRMC\\
		\hline
		I & 0.08 &  & - & 1.57 & 1.60 & 1.60 & 1.57 & 1.60 \\ 
		II & 0.09 &  & 1.26 & 1.23 & 1.26 & 1.26 & 1.26 & 1.26 \\ \\
		III& 0.26 &  & 3.26 & 2.62 & 3.02 & 3.02 & 2.44 & 3.02 \\ 
		IV & 0.29&&-&2.81 & 2.75 &2.75 &2.39&2.75\\\\
		V & 0.65 &  & 1.28 & 1.06 & 1.20 & 1.29 & 0.62 & 1.14 \\ 
		VI & 0.68 &  & - & 1.99 & 0.54 & 3.42 & 0.67& 0.30\\ 
		\hline
	\end{tabular}
	\bigskip
\end{table}

\section{Concluding Remarks\label{sec:conclusion}}
We have proposed a new imputation method for survey sampling by assuming a low-rank structure and incorporating fully observed auxiliary information. Asymptotic properties of the proposed method are investigated. One advantage of the proposed method is that we can impute the whole survey questionnaire at the same time. A simulation study demonstrates that the proposed method is more accurate than some commonly used alternatives, including inverse probability method and multiple imputation, for estimating all items. 

Our framework can also be extended in the following directions.
First, we have considered missingness at random; however, in some situations, the missingness of  $y_{ij}$ may depend on its own value, leading to missingness not at random \citep{rubin1976inference}; that is, 
$y_{ij}$ is also involved in the response probability (\ref{eqn:logit}). In this case, we will consider the instrumental variable approach \citep*{WangshaoKim2014,yang2018identification} or stringent parametric model assumptions (\citealp*{tang2003analysis}; \citealp{chang2008using}; \citealp{kim2011semiparametric}) for identification and estimation.  
Second, even though we have proposed an efficient estimator using matrix completion and derived the asymptotic bounds, its asymptotic distribution is not completely developed, which will be our future work.
Third, 
because causal inference of treatment effects can be viewed as a missing data problem, it is intriguing to develop matrix completion to deal with a partially observed confounder matrix, which is ubiquitous in practice but has received little attention in the literature \citep{yang2018identification}.

\bibliographystyle{dcu}
\bibliography{Survey_MC_abbr,ci_abbr,MC_ZL_abbr}

\end{document}